\documentclass[%
 reprint,
 amsmath,amssymb,
 aps,
]{revtex4-1}
\usepackage{multirow}
\usepackage{graphics}
\usepackage{graphicx}
\usepackage{dcolumn}
\usepackage{bm}

\usepackage[usenames,dvipsnames]{xcolor}

\usepackage{color}
\usepackage{ulem}
\definecolor{darkgreen}{rgb}{0,0.6,0}
\definecolor{darkblue}{rgb}{0,0,0.6}
\definecolor{darkred}{rgb}{0.6,0,0}
\definecolor{darkpurple}{rgb}{0.5,0,0.5}

\usepackage{hyperref}

\hypersetup{
bookmarksopen=true,
pdftitle=NirvanaCaballero-DegradationOfDomains,
pdfauthor=NirvanaCaballero, 
pdftoolbar=false,           
pdfstartview={FitH},		
pdfmenubar=true,			
pdfhighlight=/O,			
colorlinks=true,			
urlcolor=darkblue,
citecolor=darkblue,		    
linkcolor=darkpurple	    
}

\newcommand{\nccom}[1]{\textcolor{black}{#1}}

\begin{document}

\preprint{APS/123-QED}

\title{Degradation of domains with sequential field application}

\author{Nirvana Caballero}
\email[Corresponding author: ]{Nirvana.Caballero@unige.ch}
\affiliation{Department of Quantum Matter Physics, University of Geneva, 24 Quai Ernest-Ansermet, CH-1211 Geneva, Switzerland}

\begin{abstract}
Recent experiments show striking unexpected features when alternating square magnetic field pulses are applied to ferromagnetic samples: domains show area reduction and
domains walls change their roughness. We explain these phenomena with a simple scalar-field model, using a numerical protocol that mimics the experimental one. For a bubble and a stripe domain, we reproduce the experimental findings: The domains shrink by a combination of linear and exponential behavior. We also reproduce the roughness exponents found in the experiments. Our results suggest that the observed effects are due to a change in the disorder correlation length when the domain walls are subject to alternating fields during the first cycles, where the initial state of the interface plays a crucial role. Finally, our simulations explain the area loss by the interplay between disorder effects and effective fields induced by the local domain curvature.
\end{abstract}

\maketitle

\section{\label{sec:Intro}Introduction}

Ferromagnetic domains are extensively used as memory units to store information~\cite{allwood2005magnetic}, from the initial emergence of magnetic-bubble-based devices~\cite{eschenfelder2012magnetic,Malozemoff}, to more recent breakthroughs in racetrack memories based on domain wall and magnetic skyrmions~\cite{luo2020current}. Controlling such domains and understanding their behaviour is thus of great applied as well as fundamental interest. A synergistic interplay between basic research and technological development has led to advanced knowledge of how to efficiently create and manipulate magnetic domains~\cite{hayashi2008current,parkin2008magnetic}, in particular with fixed and alternating magnetic fields~\cite{bobeck1975magnetic}. However, several key aspects of domain behaviour remain less well understood, and from a basic physics view point, these objects still present striking features which appear difficult to elucidate.

In particular, the reaction of ferromagnetic domains to sequential application of alternating magnetic fields has not been extensively explored~\cite{domenichini2019transient}. Recent experiments in ultra-thin ferromagnetic films with perpendicular magnetic anisotropy revealed an unexpected domain dynamic response. Under the application of alternating magnetic square-field pulses in the creep regime, polar magneto-optical Kerr effect microscopy images showed that initially circular domains evolved towards distorted domains with irregular shape and concomitant domain area reduction~\cite{domenichini2019transient}. The causes of domain area reduction are still an open issue, as well as the change in the geometrical properties of domains.

\begin{figure}
\begin{center}
{\includegraphics[width=1\linewidth]{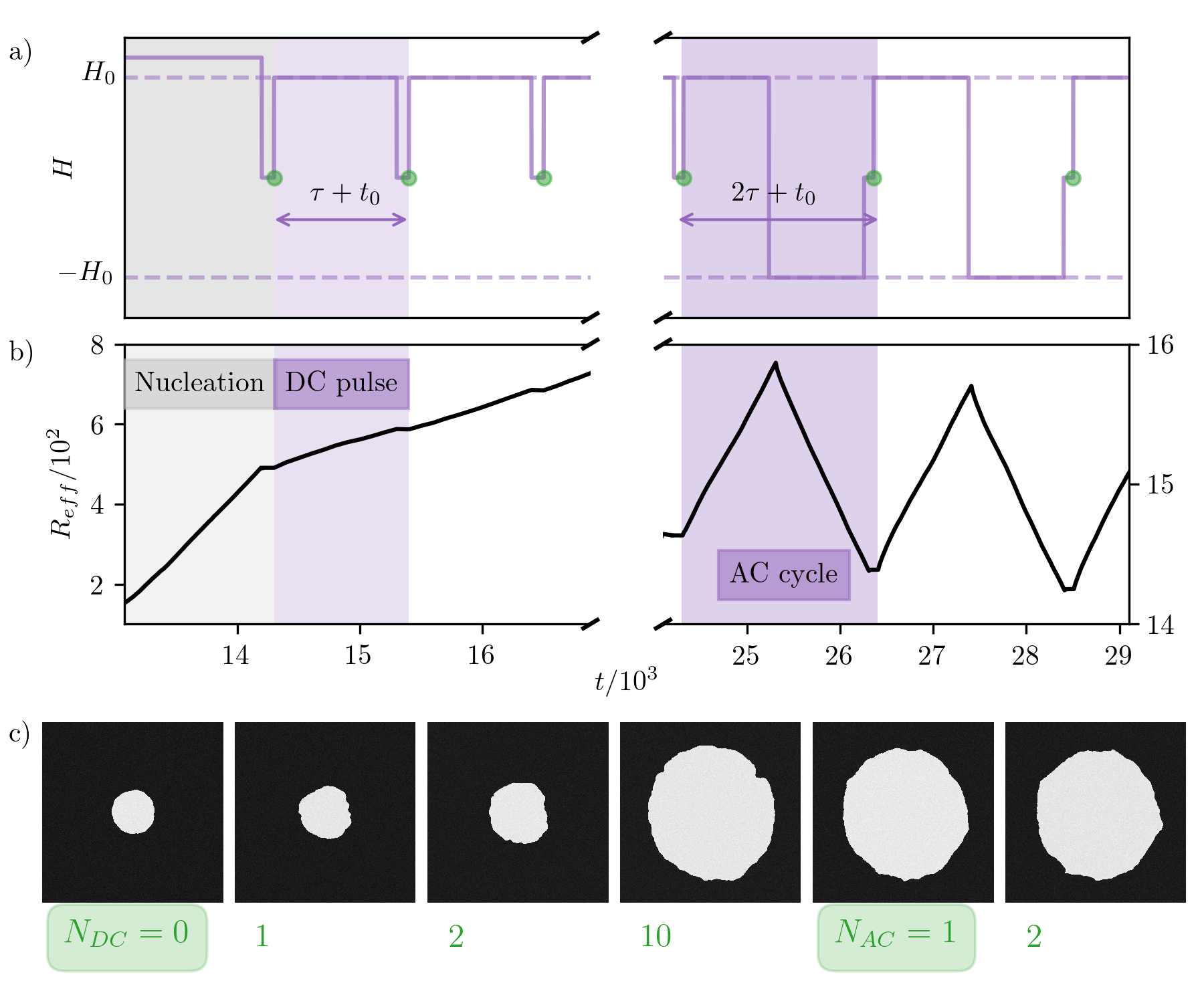}}
\end{center}
\caption{\textit{Protocol to study domain walls statics and dynamics numerically}: We emulate typical protocols used in polar magneto-optic Kerr effect microscopy. First, we imitate the nucleation process by creating a domain of radius $R$. We apply a positive field to let the domain grow until it reaches a larger effective radius. After letting the system evolve at zero field for a time $t_0$ (relaxation process), we apply $N_{DC}$ \nccom{pulses} followed by $N_{AC}$ cycles. In a) the applied field is shown. In b) we show the effective radius of a domain as a function of time. In c) we show images of the system after each relaxation time (indicated by the green dots in a)). Each image in c) is of size $L\times L$, with $L=4096$.
}
\label{fig:pulsesprotocol}
\end{figure}

Domains show very rich phenomena under the application of fields. When simple magnetic field square pulses are applied to ferromagnetic samples, velocity-field curves are typically characterized by three dynamical regimes as a function of the field magnitude: a flow regime where the velocity grows linearly, a depinning regime with signatures of the zero-temperature behavior, and a creep regime where the velocity grows exponentially~\cite{Ferrero2013nonsteady}. In the low-field creep regime, which only emerges in presence of disorder, the ultra-slow motion of domain walls occurs through thermal activation. Activated events that involve collective reorganizations trigger the domain wall and as a result the interface shows a non-zero average velocity.

In the creep regime, the velocity is highly non-linear and displays a stretched exponential behavior as a function of the magnitude of applied field $H_0$: $\sim e^{-H_0^{-\mu}}$, where $\mu$ is a universal exponent~\cite{ferrero2020creep}. In the framework of disordered elastic systems theory, this universal exponent is directly related to the roughness exponent $\zeta$ characterizing the domain wall geometry. Both exponents are related through the equation $\mu=(d+2\zeta-2)/(2\zeta)$, where $d$ is the interface dimension~\cite{nattermann_creep_full,chauve_creep_long}. There is thus an intrinsic relation between dynamical and geometrical features of interfaces. In the case of ferromagnetic ultrathin films, $\mu\simeq1/4$~\cite{lemerle_domainwall_creep,kim2009interdimensional,Jeudy2016,caballero2017excess} and $d=1$ gives a prediction of $\zeta=2/3$, which was corroborated for different materials~\cite{lemerle_domainwall_creep,Moon2013,pardo2019common}.

Theoretical and numerical studies of effects in the creep regime are difficult to tackle due to the glassy nature of the problem~\cite{ferreroPRL2017spatiotemporal}. The universality of these phenomena allows statistical-physics minimal models to successfully capture the main features of interfaces in this regime. One great advantage of treating systems through this kind of models, is that it allows us to distinguish what are the main physical ingredients responsible for the complex observed behavior.

The simplest approach is to consider ``elastic line'' type models~\cite{ferrero2020creep}. In many cases these models describe the physics of a system very well. However, at the same time, they are only able to describe interface properties and fail to characterize important bulk features. In particular, some theoretical works based on this approach predict that the application of alternating periodic magnetic fields would produce a periodic domain wall oscillation around the initial condition. Although the response is expected to be nonlinear and hysteretic, the magnetic domain is expected to remain
unchanged after the application of an integer number of alternating magnetic field cycles~\cite{Natterman_ac_1999,nattermann2001hysteretic,glatz2003domain,nattermann2004hysteresis}.

In a different description level, but still in the category of statistical-physics minimal models, Ginzburg--Landau models have been proven useful to understand many features of domain walls, especially for non-disordered systems~\cite{RevModPhys.49.435_hohehalp,kawasaki_dynamics_1977,chaikin,stariolo2007,PerezJunquera2008,MarconiPhi42011}. A connection between a minimal Ginzburg--Landau model and the quenched Edwards--Wilkinson equation, belonging to the category of elastic line models, was recently established for disordered systems~\cite{caballero_GL-EW_2020}. This could explain why several features of domain walls, for example the creep regime, are successfully described by the more complex Ginzburg--Landau type models too~\cite{caballero2018magnetic}.

More complicated approaches, such as the ones based on the Landau--Lifshitz--Gilbert equation~\cite{Malozemoff} are also useful to describe domain wall dynamics, especially if one is interested in microscopic details and internal structure of domain walls. This kind of approach is also important to address material-specific properties. However, to describe the basic physics of a system it is often not necessary to consider the full micromagnetic description. For example, it was recently shown that scaling exponents describing Barkhausen jumps in thin ferromagnetic films of Pt/Co/Pt obtained by full micromagnetic simulations cannot be distinguished from those expected for the much simpler quenched Edwards--Wilkinson equation. Thus, commonly used simple models based on describing domain walls as elastic lines correctly capture the large-scale critical dynamics of the
system~\cite{herranen2019barkhausen}. On the other hand, this approach is computationally expensive, thus limiting the system sizes that can be studied in reasonable times. 

The great advantage of Ginzburg-Landau type models is that \nccom{domain} properties may be studied in combination with interface characteristics in a very simple way. Typical protocols used in experiments to observe ferromagnetic domains in ultrathin films may be emulated, and effects in the different dynamical regimes may be studied~\cite{caballero2018magnetic}. Moreover, this simplistic description allows us the exploration of system sizes which are comparable to experimental ones~\cite{guruciaga_caballero_2021ginzburg}. 

In this work, we propose a numerical treatment based on a Ginzburg--Landau type model to study domains under alternating fields that imitates the protocol used in the experiments of Ref~\cite{domenichini2019transient}. With a GPU-based parallel algorithm to numerically solve the equation of motion, we are able to explore domain and domain wall dynamics in extremely long simulations.

In our approach, no material specific parameters need to be defined a priori, and it allows us to study in detail how domains and domain walls behave under field application. The model consists of considering an order parameter, identified as the projection of the magnetization along the easy axis, a double-well potential favouring two preferential states for the order parameter, an external field, disorder and temperature. Under the application of a sequence of alternating fields, we observe a domain area reduction which follows a linear combination of a decreasing exponential and a linearly decreasing function. This is compatible with what was reported for domain area loss in samples of Pt/Co/Pt and Pt/[Co/Ni]/Al~\cite{domenichini2019transient}. Moreover, we obtain roughness exponents characterizing the domain walls geometry which are indistinguishable from the ones reported in the same experiments.
Our results show that the scaling behavior observed in the experiments can be explained by considering a model with very few ingredients.
In particular, more complicated interactions like for example the Dyzaloshinskii--Moriya interaction (DMI)\nccom{, as suggested in Ref.~\cite{domenichini2019transient},} are not the cause of such observed phenomena. Moreover, since Ginzburg--Landau treatments are also applicable to other ferroic materials, the current analysis can be extended to other systems with the same main basic interactions.

The article is organized as follows. We first describe the model used to study the effect of alternating fields in domains (section~\ref{sec:model}). Details of the numerical and experimental protocols are presented in section~\ref{sec:NumericalProtocol} and appendix~\ref{sec:ExperimentalProtocol}, respectively. In section~\ref{sec:SimulationDetails} we give all the pertinent simulation details. In section~\ref{sec:DomainDynamics} we numerically study the evolution of bubble and stripe domains under DC (section~\ref{sec:DCdynamics} and appendix~\ref{app:ODC}) and AC fields (section~\ref{sec:ACdynamics} \nccom{and appendix~\ref{app:flat}}). Details of these analysis are presented in~\ref{app:acfit}. In section~\ref{sec:DomainGeometry} we analyze the domain walls geometries for both bubble and stripe domains with the fitting method described in appendix~\ref{app:fitting}. In section~\ref{sec:Physics} we discuss important implications of our results. We finally present our concluding remarks and perspectives in section~\ref{sec:Conclusions}.

\section{Emulating real experiments to study domain wall dynamics numerically}
\label{sec:model}

To study the effect of alternating magnetic fields on domain geometry and dynamics, we use a modified Ginzburg--Landau model with disorder~\cite{jagla2004,jagla2005,caballero2018magnetic}.

In this model, a non-conserved order parameter $\varphi(\vec{r},t)$, describing the local state of a two-dimensional system ($\vec{r}\in \mathcal{R}^2$), is governed by the Langevin equation

\begin{equation}
\eta\frac{\partial \varphi}{\partial t}
= 
\gamma\nabla^2\varphi + (D\alpha \varphi + H) (1 - \varphi^2)  +\xi.\\ 
\label{eq:Langevin}
\end{equation}

$\xi=\xi(\vec{r},t)$ is a Gaussian white noise with zero mean and two-point correlator
\begin{equation}
\langle \xi({\vec{r}}_2,t_2)\xi({\vec{r}}_1,t_1) \rangle=2 \eta T \delta^2({\vec{r}}_2-{\vec{r}}_1)\delta(t_2-t_1).
\label{eq:thermalcorrelations}
\end{equation}

In the case we are interested in, \textit{i.e}, ferromagnetic thin films with perpendicular magnetic anisotropy, as discussed in detail in Ref.~\cite{caballero2018magnetic}, $\varphi$ represents the normalized projection of the magnetization along the easy axis. $\eta$ is a damping parameter, $T$ the temperature of the system, $H$ \nccom{is a force acting at each position $\vec{r}$ in the system, that represents} an external magnetic field favoring the +1 state, and $\alpha$ and $\gamma$ are constants proportional to the out-of-plane magnetic anisotropy and the exchange stiffness, respectively. The term $D\alpha \varphi(1-\varphi^2)$, coming from a double-well potential, is responsible of favoring two preferential values for $\varphi$ ($\pm 1$). Structural quenched disorder is introduced by considering a perturbation of the double-well potential as $D=D(\varepsilon,\vec{r})=(1+\epsilon\chi(\vec{r}))$. \nccom{Here, $\chi(\vec{r})$ is a random number at position ${\vec{r}}$ taken from a Gaussian distribution with zero mean and unit variance, whose correlations satisfy  $\langle \chi(\vec{r}_i)  \chi(\vec{r}_j)\rangle=\delta^2({\vec{r}}_i-{\vec{r}}_j)$, where ${\vec{r}}_{i,j}$ are the relative distances between the simulation cells $i$ and $j$, and $\langle{\vphantom{|}\cdots}\rangle$ denotes the average over different disorder realizations}. This perturbation may be interpreted as a site-to-site variation of the energy that $\varphi$ needs to overcome in order to change its state. We recently showed that disorder introduced in this way is compatible with the so-called random bond disorder \nccom{--and translated into a pinning force with short-range correlations acting on the interfaces in the system--}\cite{caballero_GL-EW_2020}. This disorder type was shown to describe domain walls in a large family of magnetic materials~\cite{Jeudy2016}.

Through a linear transformation, equation (\ref{eq:Langevin}) can be written in reduced units: without loss of generality and since we are interested in the scaling behavior of the problem rather than in a quantitative description, in the following, time is given in units of $\frac{\eta}{\alpha}$, space is in units of $\sqrt{ \frac{\gamma}{\alpha}}$, field is given in units of $\frac{\alpha}{\eta}$, and temperature is in units of $\frac{\sqrt{\gamma \eta}}{\alpha}$.

Numerically solving equation~(\ref{eq:Langevin}), with a protocol emulating the experimental ones typically used in polar magneto-optic Kerr effect microscopy, allows us to obtain velocity-field domain walls responses. Under the application of square constant field pulses, the walls of a bubble domain acquire a velocity with particular features as a function of the field. The domain wall velocity has the same characteristics as experimentally observed velocity-field curves in ultrathin ferromagnetic systems with strong perpendicular magnetic anisotropy~\cite{caballero2018magnetic}. In particular, at sufficiently low fields this model exhibits the so-called creep regime. Moreover, if one is interested in a quantitative comparison between experimental and numerically obtained velocity-field curves, some approaches connecting the model used here and the Landau-Lifshitz-Gilbert equation have been recently proposed~\cite{guruciaga_caballero_2021ginzburg}.

In this work, we numerically solve equation~(\ref{eq:Langevin})~\cite{caballero2018magnetic}, and explore domain dynamics in the creep regime. We follow a numerical protocol which is equivalent to an experimental one recently used in polar magneto-optic Kerr effect microscopy experiments to study the effect of alternating magnetic fields on driven domains~\cite{domenichini2019transient}, as detailed in~\ref{sec:ExperimentalProtocol}. 

To reveal the causes of domain area reduction observed in the experiments, we imitate the experimental protocol in our simulations. The method consists of a series of \nccom{pulses} of two types which are called ``DC'' and ``AC''. The DC experimental protocol is the standard one used to measure domain wall velocities~\cite{metaxas2007creep,Jeudy2016,caballero2017excess,diez2018wire}, while the AC one is applied afterwards to probe domain wall dynamics and geometry under alternating magnetic fields.

\subsection{Numerical protocol}
\label{sec:NumericalProtocol}

The numerical protocol is summarized in figure~\ref{fig:pulsesprotocol}. It consists of three parts as in the experiments: nucleation, DC \nccom{pulses} and AC cycles. We imitate the nucleation process by creating a domain of radius $R_i$: we start with a system where $\varphi(\vec{r},t=0)$ takes the value $-1$, except inside a circular region of radius $R_i$, where it is $1$. We apply a square pulse of field $H_i$ above the depinning field favouring the phase corresponding to the value $1$ (in a real PMOKE experiment this step would be equivalent to the nucleation process with a high field) and we let the system evolve at fixed temperature $T$ and disorder intensity $\varepsilon$ until the domain reaches an area $\pi R_0^2$.
After letting the system evolve at zero field for a time $t_0$ to ensure a stationary value of the domain area (relaxation process), we apply $N_{DC}$ \nccom{pulses} followed by $N_{AC}$ cycles. A DC \nccom{pulse} consists of the application of a square-pulse of field $+H_0$ during a time $\tau$ followed by a relaxation time $t_0$ at zero field. An AC cycle consists of the application of a square field pulse of field $+H_0$ during a time $\tau$ followed by a square field pulse of field $-H_0$ during a time $\tau$. The system then evolves at zero field for a time $t_0$.

\subsection{Simulation details}
\label{sec:SimulationDetails}

We consider a system of size $L\times L$ simulation cells, with $L=4096$ and periodic boundary conditions in both directions. To obtain the system evolution as a function of time, we integrate equation~(\ref{eq:Langevin}) by following a semi-implicit Euler method with integration time-step $\Delta t=0.1$ in Fourier space~\cite{jagla2004,caballero2018magnetic}.

For $\varepsilon=1$, the velocity-field response of a domain with these characteristics was shown to display a depinning field $H_d\simeq0.0598$~\cite{caballero2018magnetic}. For the nucleation-equivalent step, we define an initial circular domain of radius $R_i=100$, and we let it grow until it reaches a radius $R_0=500$ under the application of a field $H_i=0.06$ at $T=0.01$. The velocity in this case is $v(H_i=0.06,\varepsilon=1,T=0.01)=0.033$. Since we are interested in the study of the effect of alternating fields in the creep regime we consider at $T=0.01$ the response of domains under field $H_0=0.05$. The velocity of a domain under constant square field pulses at this field is $v=0.010$. To ensure a growth of the domains areas of at least $\sim 10\%$ with the application of one square field pulse, we chose $\tau=10^4$. Note that each AC cycle at this field corresponds then to $2.1\times 10^5$ simulation time steps.

\section{Domain dynamics}
\label{sec:DomainDynamics}

We measure the domain area $A$ defined as the number of simulation cells for which the order parameter has a positive value. The evolution of the effective radius $R_{eff}=\sqrt{A/\pi}$ of a bubble domain is shown in figure~\ref{fig:pulsesprotocol} b). Images taken after all relaxation processes imitating the way in which images are typically obtained experimentally are also shown. 
The advantage of performing this kind of simulations is that we now have evolution details that are not easily accessible experimentally. In particular, we can precisely track the domain area evolution even in presence of an external field. By tracking the evolution of $R_{eff}$ as a function of time, many interesting features of the domain area evolution arise. First, as was already observed before~\cite{caballero2018magnetic}, immediately after removing the field the domain area is slightly increased, highlighting the importance of taking into account a relaxation time $t_0$ in the simulations. Second and more important, the velocity at which the domain area is reduced when applying a negative field, revealed by the slope of $R_{eff}$, is larger than the velocity at which the domain grows when a positive field is applied. This point raises intriguing questions about field driven domain dynamics which we address in the following sections.

\subsection{DC dynamics \label{sec:DCdynamics}}

\begin{figure*}[t!]
\begin{center}
{\includegraphics[width=1\linewidth]{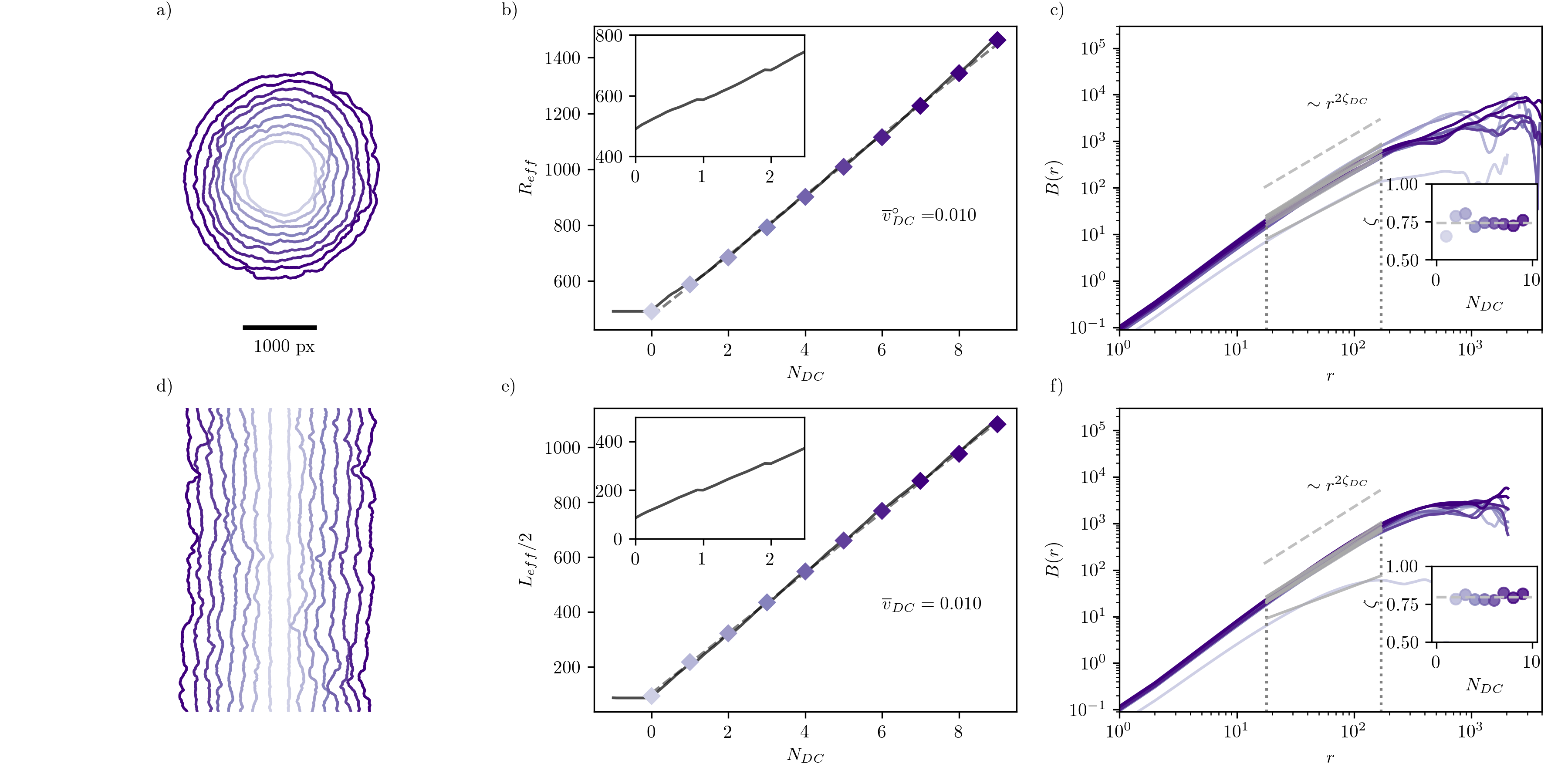}}
\end{center}
\caption{\textit{Domain expansion under the application of square field pulses (DC \nccom{pulses})}: Under the application of DC \nccom{pulses} a domain grows, and its velocity and domain wall geometry (characterized by the roughness exponent) are independent of the global domain geometry. We compute the area $A$ of domains as a function of time, from which we extract the effective radius ($R_{eff}=\sqrt{A/\pi}$) in the case of the bubble, or the effective width in the case of the stripe ($L_{eff}=A/L$) (gray line in b) and e), shown in detail in the insets). After each relaxation time in one DC \nccom{pulse} (diamonds in b) and e)), imitating the experiments, we extract the domain wall position when no external field is applied (shown in a) and d)). A fitting of this evolving domain wall position (shown in dashed gray line in b) and e)) gives a velocity for the domain walls which is independent from the domain geometry. We compute the roughness function $B(r)$ of each of the obtained relaxed interfaces (c) and f), averaged over both domain walls in the case of the stripe). These curves display a power-law behavior $B(r)\sim r^{2\zeta}$. The roughness exponents obtained by fitting the region $[17,169]$ are shown. We fit each curve in this range (the curves obtained with the fitting procedure are shown in solid gray lines in c) and f)). From each fit we obtain a roughness exponent shown in the inset of the figures on the right. For this case the average of these values (dashed gray line in the insets of c) and f)) is $\zeta^{\circ}_{DC}=0.74\pm0.04$ for the bubble domain and $\zeta_{DC}=0.80\pm0.02$ for the stripe domain.}
\label{figure2_dcdata_stripe}
\end{figure*}

The observation of larger velocities when a circular domain contracts, compared to the ones observed for domain expansions, raises one important question: What is the role of the circular shape of the domain in the dynamics? To answer this question we first focus on the more traditional DC approach, by comparing how a circular and a stripe domains grow and shrink.

As described in section~\ref{sec:model}, our numerical nucleation-like process consists of starting with an initial configuration where the order parameter takes the value -1, except for a region $\Omega$ where it takes the value 1. To analyze the evolution of a circular domain, we start with $\Omega$ defined as a region of radius $R_i=100$ centered in the middle of the system. As a first test, we let the system evolve at zero field, and no significant changes in the domain area are observed, \textit{i.e.}, even for this ``small'' circular domain the force due to the curvature is not enough to contract it in presence of disorder. This feature of circular domains is a consequence of disorder: without disorder and no applied external field, a circular domain will shrink due to the effective field sensed by the wall as a result of the domain curvature (which is proportional to the inverse of the domain radius), as discussed later in section~\ref{sec:ACdynamics}.

Another possibility is to consider a stripe domain. In this case $\Omega$ is given by a rectangular region of size $L \times L_i$. We repeat the same protocol for this case, by choosing $L_i$ and $L_0$ so that the initial and final domains in the nucleation-like numerical step have the same area compared to the circular case. We then apply $N_{DC}=10$ \nccom{pulses} to both systems and we compute the domain area in both cases as a function of time, as shown in figure~\ref{figure2_dcdata_stripe}. To emulate the way in which velocities are obtained in the experiments, we select one area point per DC \nccom{pulse} (the last one of each cycle, after the relaxation time). We fit these points with a linear function, and the slope of this fit gives the domain wall velocity. For both domain geometries, the computed velocities during DC \nccom{pulses} are the same within numerical errors, and give $\overline{v}_{DC}=0.010$. In figure~\ref{figure2_dcdata_stripe} b) and e), it can be seen that the fitting curve of these points has a barely distinguishable slope, compared to the slope of the function describing the effective radius or length of the domains as a function of time.

To study the effect of negative fields, we follow a DC protocol for a field $-H_0$. We call to this protocol ODC, where the ``O'' stands for opposite. In this case, we obtain a shrinking velocity $\overline{v}^{\circ}_{ODC}=-0.013$ and $\overline{v}_{ODC}=-0.011$ for the bubble and stripe domains respectively, as shown in~\ref{fig:ODC}. The observation of larger velocities for shrinking domains motivates a deeper analysis that we address in the next section.

\subsection{AC dynamics \label{sec:ACdynamics}}

\begin{figure*}[t!]
\begin{center}
{\includegraphics[width=1\linewidth]{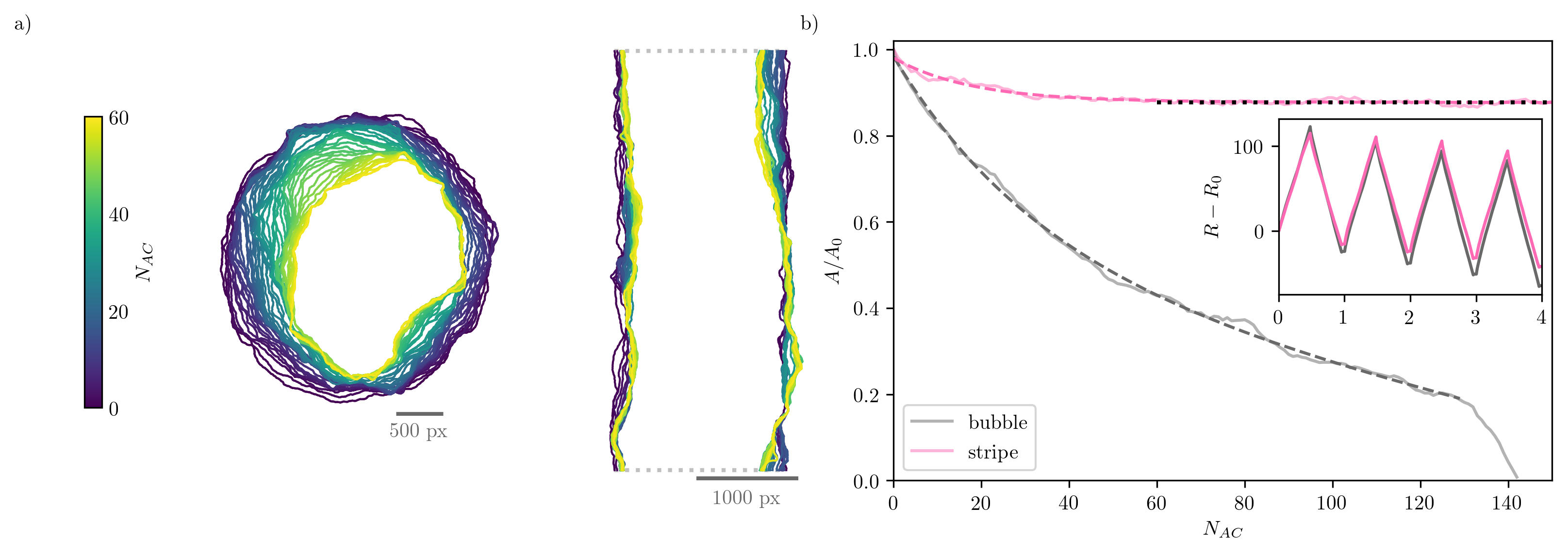}}
\end{center}
\caption{\textit{Area loss under the application of alternating fields}: An AC cycle consists of the application of two square field pulses of the same intensity and duration, but with opposite sign, followed by a relaxation time. A bubble and a stripe domain under the same simulation conditions show area loss during this process. In a) the domain boundaries after $N_{AC}$ cycles are shown. In b) the domain area loss computed with respect of the domain initial area $A_0$ is also shown as a function of the number of AC cycles for the bubble and stripe. The dashed lines are fittings of the curves with a linear combination of an exponential decay and a linearly decreasing function. The bubble domain collapses after 143 cycles. The stripe domain also shows area loss, but to a less drastic degree than the bubble domain \nccom{and after around 45 $N_{AC}$ cycles it fluctuates around its mean value, shown in black dotted lines}. Details of the domain area loss with respect to the initial condition for the bubble and the stripe domains are shown for the first 4 cycles in the inset of b).
}
\label{fig:areas}
\end{figure*}

The observation of larger velocities for shrinking domains compared to velocities of expanding domains motivates a deeper study of the situation. In this section, we analyze in detail the evolution of domain area under the AC numerical protocol selected to emulate the experimental one.

Starting with a circular and a stripe domain which were subjected to the previously described DC \nccom{pulses}, we analyze their area under the application of multiple AC cycles~\footnote{One more pulse of field $+H_0$ was applied to the stripe domain. The duration of the pulse was defined in order to allow the stripe domain to reach the same value of the initial area in the bubble case. This process was followed by a relaxation time $t_0$.}. Both systems are subject to the same conditions (\textit{i.e.} both simulations are performed with exactly the same parameters, and the only difference is given by the initial condition for the order parameter).

As shown in figure~\ref{fig:areas}, both systems show area loss during this process. In particular, the bubble domain completely collapses after 143 AC cycles, while the stripe domain also shows area loss, although to a less drastic degree. In the experiments shown in Ref.~\cite{domenichini2019transient}, the area loss is described by a linear combination of an exponential decay and a linearly decreasing function. In the case of our simulations, the curves are also very well fitted by such a combination for the bubble (over a wide range) and the stripe domains, as indicated with dashed lines in figure~\ref{fig:areas}. A detailed inspection of the area evolution reveals that in both cases the expansion velocity of the domains is larger than the velocity acquired by a contracting domain. Surprisingly, and to the best of our knowledge, this is something which is not usually tested in experiments. However, the observations presented in~\cite{domenichini2019transient} are compatible with this claim.

\begin{figure*}
\begin{center}
{\includegraphics[width=1\linewidth]{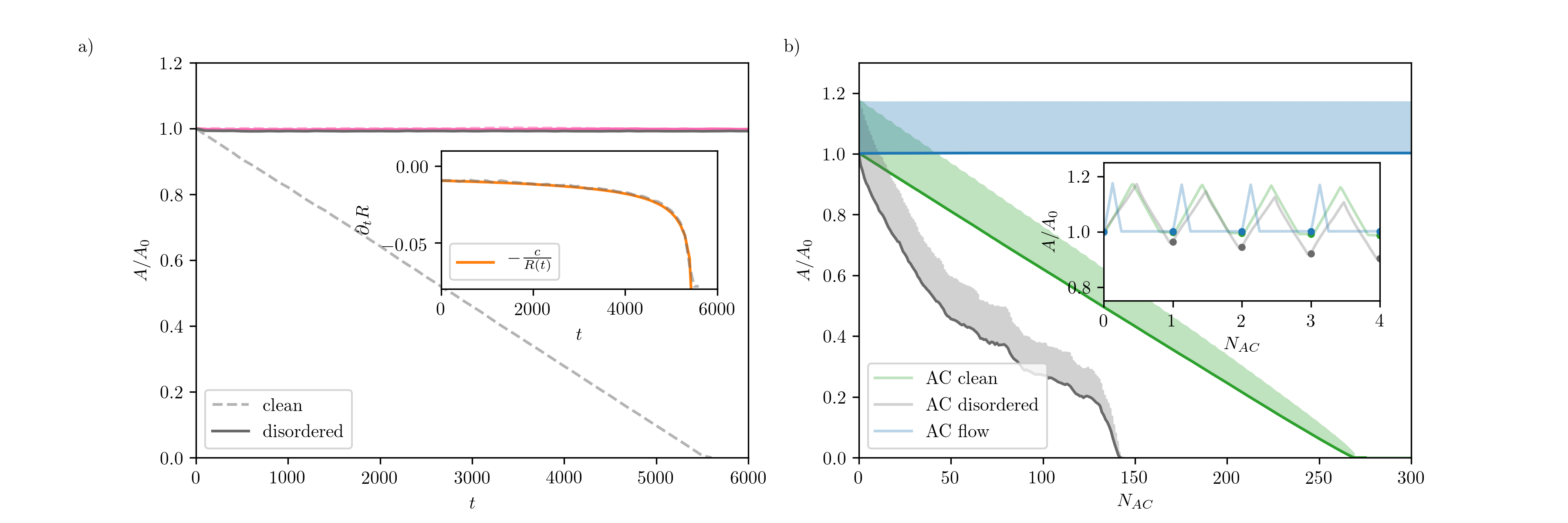}}
\end{center}
\caption{\textit{Stability of domains}: In a) we study the evolution of domain areas with respect to its initial value $A/A_0$ for a clean (dashed lines) and a disordered system (continuous lines). We consider a bubble (gray) and a stripe (pink) domain under zero applied field. The only case in which a domain losses area is the bubble domain in a clean system. This effect is due to a field $-c/R$ (with $c$ the energy of the domain wall and $R$ the domain radius) induced by the domain curvature. In the inset of a) we show how the domain radius variation as a function of time $\partial_t R$ is indeed described by $-c/R(t)$. In b) we show details of bubble domains areas under AC fields for a disordered (gray) and a clean (green) system at field $H_0=0.05$. In blue, we show the area evolution of a bubble domain under AC cycles for a field $H_0=0.5$. In the inset of b) we show a close up for the first 4 $N_{AC}$ cycles of these three studied cases. For these cases the time pulse duration $\tau$ was chosen in order to let a domain grow from an initial radius 1500 to the same final radius.}
\label{fig:relax}
\end{figure*}

To further explore the causes of domain area loss under AC cycles we exploit many different scenarios. First, we recall that the domain wall curvature induces a field $-\frac{c}{R}$, where $R$ is the domain radius and $c$ the domain wall energy~\cite{caballero2018magnetic}. In the case of a perfect stripe domain this effect is negligible since $R\to\infty$. Under zero applied field a stripe domain will exhibit a constant area for clean ($\varepsilon=0$ in equation (\ref{eq:Langevin})) and for disordered systems as well ($\varepsilon \neq 0$) (see figure~\ref{fig:relax} a)). The case of the bubble domain is different: the field proportional to its curvature is not always negligible. In presence of disorder this effect might be hidden and the area of a bubble domain will remain constant. It is indeed in this principle in which typical Polar-magneto optic Kerr effect microscopy experiments rely: images are taken at zero field with a quasi-static technique in which the structure of the domains remains stable during image acquisition~\cite{domenichini2019transient}. In other kind of systems like ferroelectrics, it was also shown that pinning of domain walls plays a key role in stabilizing domains~\cite{blaser_domainsizestabilityferroelectrics_2012}.

We check that for the parameters chosen in this work a circular domain will not shrink only due to its curvature. We perform simulations at zero field for a clean and a disordered system ($\varepsilon=1$) for a stripe and a bubble domain and compute its relative areas as a function of time. As can be seen in figure~\ref{fig:relax} a), the only case in which a domain losses area under zero field is a bubble domain in a clean system. This instability of a domain in a clean system is further confirmed by micromagnetic simulations in clean systems where the interfacial DMI coefficient is negligible~\cite{wang2018theory}. Moreover, since in the system considered in this work the domain wall energy \nccom{can} be easily computed~\cite{caballero_GL-EW_2020} and gives $c=2\sqrt{2}/3$, we check that the variation of the domain radius as a function of time is compatible with an effective applied field $-c/R$, as shown in the inset of figure~\ref{fig:relax} a).

The area loss due to curvature effects can be further explored by applying AC cycles to a bubble domain in a clean system. To fairly compare the evolution of a domain in a clean system under AC cycles, we define the time duration of the pulse of an AC cycle $\tau_{\textrm{clean}}=1860$. This value is chosen in order to let a bubble of initial radius $R_0\simeq 1500$ reach the same final area after a square-field pulse of field $H_0=0.05$ compared to one evolving in a disordered system, for which we recall, we chose $\tau=10^4$ (see section~\ref{sec:SimulationDetails}). As can be seen in figure~\ref{fig:relax} b) under this protocol a bubble domain will still lose area after each AC cycle, but the area loss will follow a linear function as a function of the number of $N_{AC}$ cycles, and as a consequence a bubble domain will collapse after a larger number of cycles.

If instead we choose a field $H_0$ large enough so the field felt by the domain wall due to curvature effects is negligible, then the domain area should remain unchanged. We verify this claim by studying the area evolution of a bubble under an AC protocol with a large field $H_0=0.5$, where the velocity corresponds to the flow regime (and we chose $\tau_{\textrm{flow}}=170.65$~\footnote{To have enough time resolution in this case we integrate the equation of motion~(\ref{eq:Langevin}) with an integration time-step equal to $10^{-2}$. All other simulations details remain unchanged.} determined as in the AC clean protocol). We performed 1000 AC cycles under this condition and the domain area remains constant after each AC cycle, as can be seen in figure~\ref{fig:relax} b) for the first 300 cycles, confirming the physical picture of domain area loss due to curvature effects.

Larger velocities for shrinking bubble domains are then the result of the effects \nccom{of disorder and} of the effective field felt by the interface due to its curvature. When a bubble domain is expanded the contribution of this field is reduced as the domain grows. When a bubble domain is shrinking this field becomes larger and thus the effective field driving domain collapse is increased. Disorder also causes a more rapid domain area loss in the bubble case compared to a bubble in a clean system (see figure~\ref{fig:relax} b)). \nccom{However, the fact that we observe area reduction for stripe domains shows that the area loss is not only a consequence of the domain curvature. In the stripe case, we observe an area reduction of around 10 percent in the first $\sim 45$ AC cycles. After these first AC cycles the stripe domain area fluctuates around its mean value as shown in figure~\ref{fig:areas}. The only broken symmetry for the stripe case is the order in which the field pulses are applied, and thus, the initial condition from which the domain starts its dynamics.}

\nccom{We further confirm this picture by initializing a stripe domain in a completely flat state and apply 45 AC cycles in different orders ($\Lambda$ and V, where $\Lambda$ is an AC cycle as shown in figure~\ref{fig:pulsesprotocol}, and V is one where the first pulse of the AC cycle is $-H_0$ followed by a pulse $H_0$). Depending on the order in which the pulses are applied we observe area loss or area gain, as shown in~\ref{app:flat}. This means that the geometry of the wall plays a crucial role in the area change effect (because after some AC cycles the effect in the area change, if present, is negligible independently of the order of the pulses).}
By inspecting the fitting of the curves (shown on~\ref{app:acfit}) we can conclude that an interplay between domain curvatures and disorder is responsible for the scaling behavior. There is then an intrinsic relation between domain walls geometry and dynamics.

\section{Domain wall geometry}
\label{sec:DomainGeometry}

\begin{figure}
\begin{center}
{\includegraphics[width=1\linewidth]{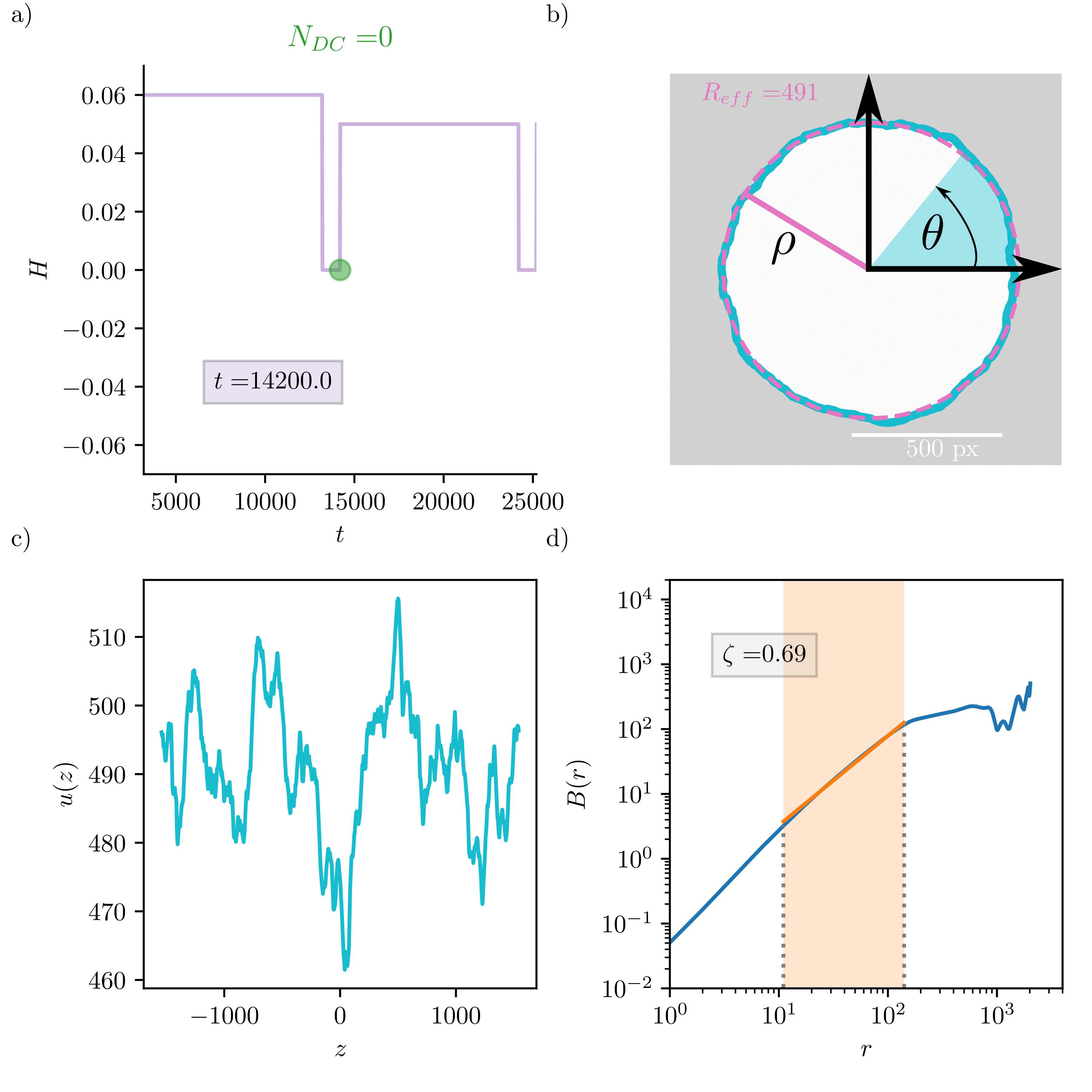}}
\end{center}
\caption{\textit{Domain wall geometry analysis}: The boundary of the domain is established (cyan curve in b)) and its difference, $u(z)$ shown in c), with respect to a perfect circle (dashed pink line in b)) which has the area of the domain, and is centered in the centroid of the domain, is computed. $z=\rho\theta$ is a standard linear coordinate defined in terms of polar coordinates. The roughness $B(r)$, defined as the average of the quadratic correlations of $u(z)$, is fitted in a restricted range (shown in orange in d)) with a power-law $\sim r^{2\zeta}$, which gives the roughness exponent $\zeta$.
The images correspond to time $14.2\times 10^3$, after the nucleation-like numerical process a).
}

\label{fig:nucleation}
\end{figure}

\begin{figure*}[!t]
\begin{center}
{\includegraphics[width=1\linewidth]{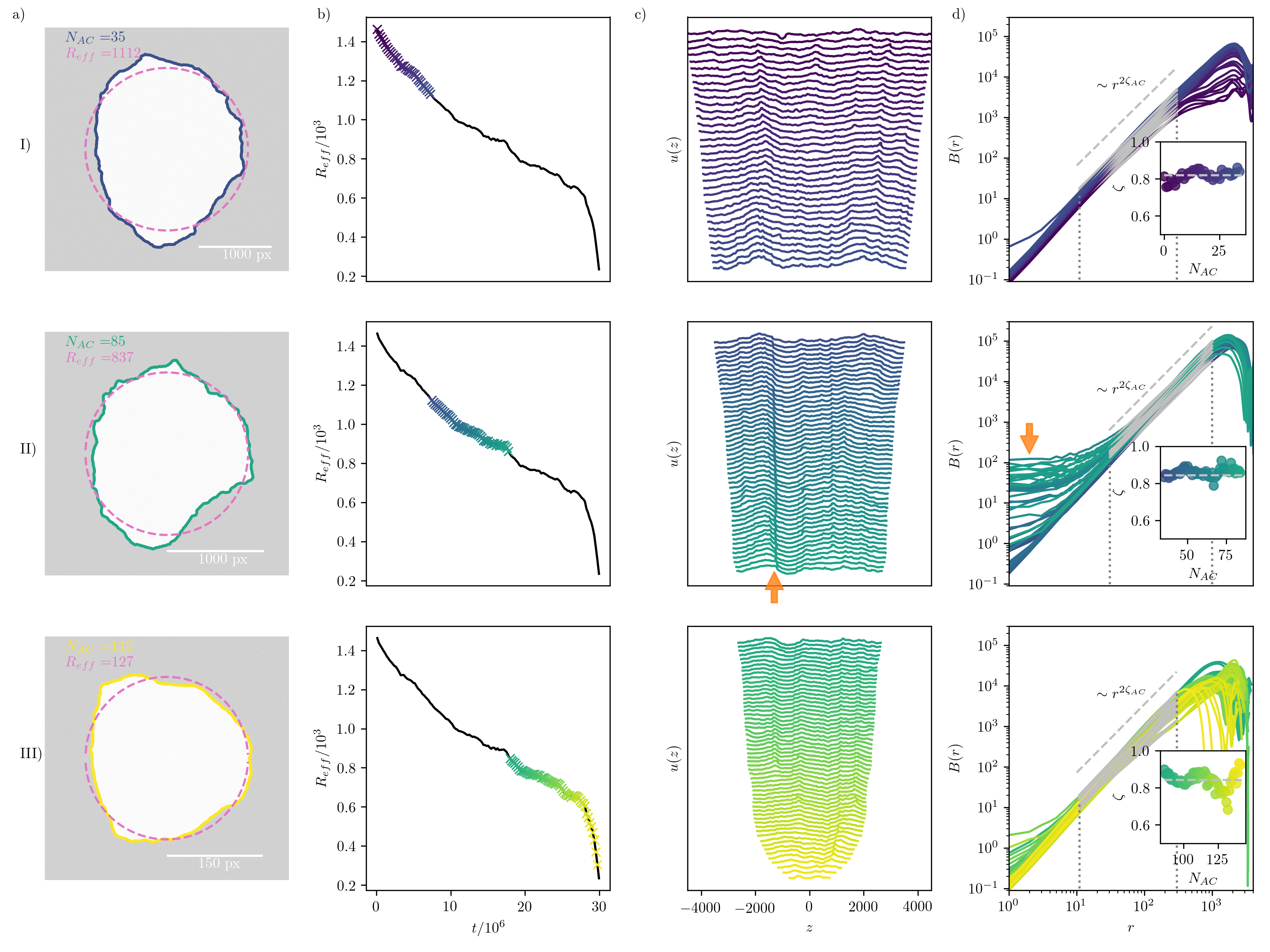}}
\end{center}
\caption{\textit{Geometry of interfaces of a bubble domain during AC cycles}. We separate the interface evolution under AC cycles in three cases: cycles 1-35 (row I)), cycles 36-85 (row II)), cycles 86-142 (row III)). In column a) snapshots of the system configuration are shown. In these snapshots the detected interface, and a circle of equivalent area to the domain, centered on the domain centroid (pink dashed line) are also shown. Column b) shows the domain area evolution, and highlighted with crosses are the points analyzed on the row. Column c) shows $u(z)$, defined as the fluctuations of the domain wall with respect to a perfect circle with the same area. Column d) shows the roughness function obtained for each $u(z)$, represented with the same color. A fit of these curves in the region indicated by dotted lines is shown in gray for each curve. The fitting function is a power-law $\sim r^{2\zeta}$. The obtained $\zeta$ values are shown in the insets. The dashed gray line in the insets corresponds to $\zeta_{AC}$, the average value of $\zeta$ over the 142 analyzed AC cycles. The dashed gray lines on the main plots in column d) are proportional to $r^{2\zeta_{AC}}$. In row II), the interface is highly pinned (see the protuberance indicated by the orange arrow in plot II-c). This protuberance induces an increase of $B(r)$ at short distances (also indicated by an orange arrow in plot II-d), and a shift of the fitting region. This motivated the division of the analysis in three parts.
After 143 $N_{AC}$ total cycles, the system is in the saturated state and no domain is observed. The roughness exponent, defined as the mean value of the roughness exponent obtained for each individual $B(r)$, is $\zeta_{AC}=0.84\pm 0.04$.}
\label{fig:acdata}
\end{figure*}

The roughness function, defined as 
\begin{equation}
B(r=|z_2-z_1|)=\langle [u(z_2)-u(z_1)]^2\rangle
\label{eq:B(r)}
\end{equation}
measures the quadratic correlations of $u(z)$, the position of the interface of a domain for a given coordinate $z$. In (\ref{eq:B(r)}) $\langle\cdots\rangle$ is an average over all possible values $z_1$ and $z_2$ for which $r=|z_2-z_1|$. This observable is widely used to study diverse systems with domain walls~\cite{barabasi,santucci_PRE_2007_fracture,paruch2013nanoscale,Jordan2020,rapin_2021_roughness}.

Usually the roughness function presents power-law behaviors with signatures of the underlying physics of the system. In particular, at short length-scales, $B(r)$ is expected to show a thermal regime. In this regime $B(r)$ follows a power-law $\sim r^{2\zeta_{th}}$, with $\zeta_{th}=1/2$. At larger scales, disorder plays a key role inducing a different power law-scaling $\sim r^{2\zeta}$~\cite{agoritsas_review,agoritsas2013static,caballero_GL-EW_2020}. The roughness exponent $\zeta$ is used to characterize domain wall geometries, which are usually determined by the competition between the few key ingredients that govern the system. Between the short-scale power-law regime due to thermal fluctuations and the large-scale power-law regime dominated by disorder a third regime \nccom{is present. This intermediate ``excess'' regime is the result of the interplay between temperature and disorder~\cite{agoritsas2013static}. The excess roughness due to disorder, compared to bare thermal fluctuations, is described at short length-scales (lower than $r_0$) by a power-law $\sim r^{\zeta_{dis}}$. We find $\zeta_{dis}\simeq 0.91$ and $r_0$ proportional to the disorder correlation length felt by the interface~\cite{caballero_crossover}}. A detailed
discussion about this is included in a forthcoming paper.

When no fields are applied, domain walls described by the model given by (\ref{eq:Langevin}) display, at low-temperatures, the features of an interface described by the Edwards--Wilkinson~\cite{edwards_wilkinson} (EW) universality class~\cite{caballero_GL-EW_2020}. However, the geometry of domain walls when non-zero fields are considered, to the best of our knowledge, has not yet been studied for this model. Moreover, in the same experiments with ferromagnetic samples where the implementation of the DC-AC dynamic protocols was established, the roughness exponents of domain walls were reported~\cite{domenichini2019transient}. In this work we compute the roughness exponents observed for the simulated domain walls and we compare our results with the experiments.
To do so we define the domain wall as the contour in which the order parameter $\varphi$ is equal to zero. As an example, a detected domain wall is shown in figure~\ref{fig:nucleation} for the bubble domain case. To study the domain wall geometry, and as was proposed to analyze the experimental domain walls in Ref.~\cite{domenichini2019transient}, we define a circle of radius $R_{eff}=\sqrt{A/\pi}$, where $A$ is the domain area, centered in the centroid of the domain. We define the function $u(z)$ as the difference between the domain wall boundary and this circle. An obtained interface $u(z)$ is shown in figure~\ref{fig:nucleation}, after the numerical nucleation-like process for the same system studied in the previous sections. The roughness function, defined in (\ref{eq:B(r)}), was computed for this interface and fitted in a region where it shows a power-law behavior. In this work we focus on the large-scale power-law scaling of the roughness since we are interested in a comparison with the experimental observations. How to choose the fitting region for a given roughness function $B(r)$ is not a trivial task (the reader might refer to~\cite{Jordan2020} for a detailed discussion of this issue in the analysis of experimental interfaces in ferromagnetic systems). To overcome this issue, in this work we analyze in detail many possible fitting regions $[r_i,r_f]$, and compute the roughness exponents and its uncertainties by following the method discussed in detail in~\ref{app:fitting}. It consists of repeating the analysis for multiple values $r_i$ and $r_f$ and analyzing the goodness of the fit for each region.

As an example, we show a fitting of a domain wall obtained after the nucleation-like process in the region $r=[10,150]$ in figure~\ref{fig:nucleation}. For this particular case we obtain a roughness exponent $\zeta=0.69$, which is remarkably close to the value $2/3$, corresponding to the `random-bond' regime for the \nccom{one-dimensional} EW \nccom{--or Kardar-Parisi-Zhang~\cite{kardar_1986_originalKPZ_PhysRevLett56_889}--} universality class.

\begin{table}[h!]
\centering
\begin{tabular}{l|lccc} 
 \hline
& System & $\zeta_{DC}$ & $\zeta_{AC}$ & $\zeta_{ODC}$\\
\hline
\multirow{2}{4.8em}{Simulation} & Bubble  & $0.74\pm0.04$ & $0.86\pm0.05$ & $0.83\pm0.06$\\ 
& Stripe & $0.80\pm0.08$ &  $0.87\pm0.03$ &$0.79\pm0.04$\\
\multirow{2}{4.8em}{Experiment} & Pt/Co/Pt & $0.73\pm0.04$ &  $0.79\pm0.03$\\
& Pt/[Co/Ni]$_4$/Al & $0.64\pm0.05$ & \\
\hline
 \hline
\end{tabular}
\caption{Roughness exponents obtained in simulations of bubble and stripe domains under DC and AC dynamics in this work, compared with the values reported for experiments in ferromagnetic samples under equivalent conditions in Ref.~\cite{domenichini2019transient}. Exponents obtained under a DC protocol but with field $-H_0$ are also shown (ODC protocol).}
\label{table:zetas}
\end{table}

In figure~\ref{figure2_dcdata_stripe}, we show the roughness functions obtained after each DC \nccom{pulse} for the bubble and stripe domains. In the case of the stripe domain, the function $u(z)$ is defined as all the points where the order parameter $\varphi$ is equal to zero. After each DC \nccom{pulse}, $B(r)$ is independently computed for the left and right domain walls, and an average over these two functions is taken to compute the roughness function after each cycle. By fitting these functions in the region  $r=[17,169]$, selected with the method described in~\ref{app:fitting}, we obtain an average roughness exponent for the bubble and the stripe domains which are indistinguishable with each other. These roughness exponents are in excellent agreement with the exponents observed experimentally, as summarized in Table~\ref{table:zetas}.  

In figure~\ref{fig:acdata} we show the interfaces obtained for the bubble domain at the end of each AC cycle, and its corresponding roughness function. These functions show very interesting features. First, during the first 35 AC cycles, besides the domain area reduction already discussed in section~\ref{sec:ACdynamics}, which is translated in this context to interfaces $u(z)$ which become shorter at the end of each AC cycle, the roughness functions display similar power-laws in the region $[10,300]$. Only a global increase of the value of $B(r)$ is observed after each AC cycle. The subsequent roughness functions (corresponding to cycles 36 to 85) show a high increase at very short distances exactly when the interface seems to be highly pinned. This strong pinning center shifts the region where $B(r)$ can be reasonably fitted with a power-law to larger values of $r$, $[30,1000]$. When the interface is depinned from this strong pinning center, $B(r)$ again recovers the behavior observed during the first 35 cycles. The power-law behavior is observed in the region $[10,300]$, but this time, the global value of $B(r)$ decreases after each AC cycle. 

In the stripe domain case, the roughness functions obtained after each cycle are similar between each other, and only a global increase on $B(r)$ with the number of AC cycles is observed. In this case the mean roughness exponent is similar to the one obtained for the bubble domain case as shown in Table~\ref{table:zetas} and discussed in~\ref{app:fitting}. We repeat the analysis for the case of ODC \nccom{pulses} (DC \nccom{pulses} with negative fields). Remarkably, the obtained roughness exponent of the stripe case under the DC protocol coincides with the one obtained for the ODC protocol. However, for the bubble case $\zeta_{ODC}$ \nccom{is} closer to $\zeta_{AC}$. This further reinforces the close relation between dynamic and static scaling exponents: while the velocity under DC and ODC \nccom{pulses} is very similar for the stripe case, also the roughness exponents $\zeta_{DC}$ and $\zeta_{ODC}$ are very close to each other. In the case of the bubble domain, where the velocity under the ODC protocol is larger than the one under a DC protocol, the roughness exponents are also different.

\section{Discussion}
\label{sec:Physics}
Based on our analysis, we find that a simple numerical model captures quite effectively the experimentally observed behavior of ferromagnetic domains under the effect of alternating magnetic fields.
In the experiments and in our simulations domains under \nccom{DC pulses followed by} repeated AC cycling exhibit pronounced area reduction.
In the experiments, many AC cycles can be applied to a ferromagnetic sample before the domain collapses. It is natural that in our work this effect happens at a shorter number of cycles, since the systems we are simulating are smaller than the experimental ones. We can estimate the orders of magnitude of our simulated systems by using some micromagnetic concepts. In a micromagnetic model, where the magnetization is ruled by the stochastic Landau-Lifshitz-Gilbert equation, the domain wall width when dipolar interactions are negligible is given by $\tilde \Delta=\sqrt{\frac{\tilde A_n}{\tilde K}}$~\cite{Malozemoff}, where $\tilde A_n$ is the \nccom{exchange} stiffness and $\tilde K$ the anisotropy. For the model considered in this work (\ref{eq:Langevin}), the equilibrium domain wall width is $\Delta=\sqrt{\frac{2\gamma}{\alpha}}$~\cite{caballero_GL-EW_2020}. 
By choosing our model parameters in order to recover the domain wall width predicted by the micromagnetic model, we can write the model parameters in terms of experimental quantities, and thus compare our numerical results with a specific material.

In particular, for Pt/Co/Pt where $\tilde A_n=14 pJ/m$ and $\tilde K=364 kJ/m^3$~\cite{metaxas2007creep}, the analogy shows that our simulated system has a lateral size of $\simeq 18 \mu m$, which is smaller than the domains studied experimentally in Ref.~\cite{domenichini2019transient}, where the initial domain area after nucleation is $\sim 4700 \mu m^2$ (giving an effective radius of $\sim 39 \mu m$). This difference could explain why our circular simulated domain collapses completely after a hundred $N_{AC}$ cycles, while in the experiments thousands of cycles can be studied before the domain collapses. However, the domain area reduction in the experiments and in our simulations follows the same functional behavior as a function of the number of applied AC cycles. A linear combination of an exponentially decreasing and a linearly decreasing functions fits the area reduction very well but a justification of the microscopic origin of behavior is still lacking. In this direction, our work provides valuable information to develop a theory describing these phenomena. The linear part may be attributed to the local curvature of domains that induces an effective field whose contribution becomes more important when a domain is shrinking. While, as we have shown, the exponential part dominating at small number of AC cycles is the result of \nccom{disorder effects, where the initial state of the domain wall plays a crucial role}.

The roughness exponents characterizing the geometry of domain walls are in excellent agreement with the ones reported in the experiments. 
\nccom{The roughness exponents observed for interfaces under AC cycles is larger compared to the ones observed for interfaces under DC pulses. The length-scales at which the disorder-dominant exponent $2/3$ is observed in the roughness function depend, among other factors, on the disorder correlation length~\cite{agoritsas2013static,caballero_crossover}. When the interface is subjected to AC cycles it is forced to wonder repeatedly trough the same disorder landscape. The disorder correlation length felt by the interface is thus increased, and the region where the $2/3$ exponent is observed is pushed to larger scales. An intermediate power-law regime with a larger roughness exponent is then evidenced~\cite{caballero_crossover}. This intermediate excess regime is widely explored and studied by us in a forthcoming paper.}

\nccom{Note that the picture of disorder correlation length increase under AC cycles is also compatible with the observed domain area change: the initial state from which the domain starts the AC cycles corresponds to one which is equilibrated for a different disorder landscape condition. After some AC cycles, this initial condition is forgotten and the domain area fluctuates around its mean value for the stripe domain or decreases linearly due to curvature effects in the case of the bubble domain.}

\section{Conclusions and perspectives}
\label{sec:Conclusions}

Motivated by recent experimental observations of pronounced domain area reduction under the application of alternating magnetic fields in ferromagnetic thin films~\cite{domenichini2019transient}, hitherto not predicted theoretically, we propose a numerical protocol to study the same effect in a very simple two-dimensional scalar-field model. The model only considers the main and basic ingredients of ferromagnetic systems~\cite{caballero2018magnetic}, and has the advantage of being material-independent (no physical units need to be specified \textit{a priori}). Its simplicity allows us to do extremely long simulations to study domain area loss under several combinations of magnetic pulses.

Our approach allows us to observe the same effect reported experimentally: under the application of alternating magnetic fields of equal duration and intensity but opposite direction (AC cycles), a domain area loss occurs. We study the effect of AC cycles over different domain geometries: a bubble and a stripe domains. Several repetitions of AC cycles lead to a collapse of the bubble domain. The stripe domain does not collapse in the same interval, but area loss is nonetheless observed. As had been proposed for the analysis of the experimental results, the domain area evolution for both domain geometries is very well fitted by a linear combination of an exponentially and linear decreasing functions. The slower decrease of the stripe domain area could indicate that this domains are better preserved compared to the bubble domains, depending on the fields treatments that could be applied to a ferromagnetic sample. The difference in the area reduction for the two domain types could be important to determine the domain wall surface tension, a quantity that is difficult to asses experimentally, but which has several important technological applications~\cite{PRAsurfacetension2018}. 

We also analyze the evolution of domain wall geometry under the application of sequential field pulses with the same direction (DC dynamics) and under AC dynamics. We characterize the bubble and stripe domain walls with the roughness exponent $\zeta$. Both geometries give similar results: a lower exponent value in the DC scenario, and a higher one under the application of AC fields. Remarkably, the exponent values in both cases are indistinguishable, within error bars, from the ones reported experimentally for Pt/Co/Pt and Pt/[Co/Ni]/Al. This result has important consequences. The relation between domain walls dynamics and geometry clearly needs further exploration. In this sense, our work can mark a course of how this problem can be approached theoretically. \nccom{On the other hand, our work shows that a treatment with AC cycles can give access to regimes which are usually hidden below experimental resolution.}

From the experimental point of view, our work contributes to the systematization of experiments. With the same experimental set-up usually used to study domain walls dynamics with polar magneto-optic Kerr effect microscopy, more information can be extracted from the same sample: besides of the usual velocity determination from domains expansion, our work shows that it is also important to determine the contraction velocity \nccom{and the role of the initial state of domains}. 

In addition, the result presented in this work could have important technological implications. Although we did not discuss any particular device implementation, according to our simulations the effect of area loss under the application of alternating magnetic fields should be observed in any disordered ferromagnetic material with strong easy axis of magnetization if the local domain curvature induces effective fields which are comparable to the fields use to manipulate domains. Of course, this claim should be verified experimentally and numerically for specific devices. Our work gives important insight into how to improve the design of devices where domains are the control unit.

It would be of extreme interest to understand which interactions could compensate the observed effect in order to prevent domain area loss. In this direction, the effects of long-range dipolar interactions (which can be easily included in the model used in this work, as for example proposed in Ref.~\cite{jagla2004}) could be explored. In this case, the area loss effect, if present, may be modified in the presence of more than one domain. The effects of currents will also be interesting to assess. In this case, more deep modifications need to be done to the model used in this work.

Our work triggers new directions to explore and at the same time, gives insight in how to theoretically understand the observed phenomena.

\section*{Acknowledgements}
I gratefully acknowledge Thierry Giamarchi, Jean--Pierre Eckmann, and Patrycja Paruch for their critical reading of the manuscript and their insightful comments. I also acknowledge valuable discussions with Sebastian Bustingorry, E.~Ezequiel Ferrero and Alejandro B.~Kolton about code implementation. I am grateful to Gabriela Pasquini for pointing me to her paper~\cite{domenichini2019transient}, which was the initial seed that stimulated me to carry out the present project. I am also grateful to Vincent Jeudy, Javier Curiale, Lucas J.~Albornoz, and Rebeca Diaz-Pardo for giving me insight in how PMOKE experiments are carried out. I thank the \textit{Imagerie et dynamique en magn\'etisme} (IDMAG) team of the \textit{Laboratoire de Physique des Solides} for their warm hospitality during my stay in their laboratory, in which I developed and implemented part of the code used in this work. 
I thank useful discussions with Thierry Giamarchi, Jean--Pierre Eckmann, Patrycja Paruch, Vincent Jeudy, Alejandro B.~Kolton, and Vivien Lecomte.
I acknowledge support from the Federal Commission for Scholarships for Foreign Students for the Swiss Government Excellence Scholarship (ESKAS No. 2018.0636).
This work was supported in part by the Swiss National Science Foundation under Division II.
All the simulations and analysis presented in this work were performed in the \textit{Mafalda} cluster of GPUs, at the University of Geneva.

\appendix 
\section{Experimental protocol to study domain walls statics and dynamics under alternating magnetic fields}
\label{sec:ExperimentalProtocol}

To study domain wall response under alternating magnetic fields by means of polar magneto-optic Kerr effect microscopy (PMOKE), the following sequence of steps was recently proposed~\cite{domenichini2019transient}.

\underline{DC experimental protocol:} In the experiment, the sample is i) saturated by the application of a high magnetic field perpendicular to the sample plane. ii) A magnetic field in the opposite direction is applied to nucleate a domain. iii) After nucleation, a square pulse of magnetic field $H_0$ during a time $\tau$ results in the expansion of the domain(s). iv) After this step, an image of (usually one) domain is optically captured (with no external magnetic field applied). The application of field pulses of equal magnitude and of the same duration $\tau$ is repeated $N_{DC}$ times. From the average displacement of the domain wall, the mean domain wall velocity for that particular field $H_0$ is extracted.

\underline{AC experimental protocol:} After the DC protocol, $N_{AC}$ cycles are applied to study the domain response. An AC cycle consists of the application of a square pulse of field $H_0$ during a time $\tau$, followed by a second square pulse of field $-H_0$ of the same duration. After each such cycle, an optical image is captured with no applied external field. The sequence of applied magnetic fields is shown in figure~\ref{fig:pulsesprotocol}. 
\\

\section{Domain shrinking: ODC protocol}
\label{app:ODC}

\begin{figure*}
\begin{center}
{\includegraphics[width=1\linewidth]{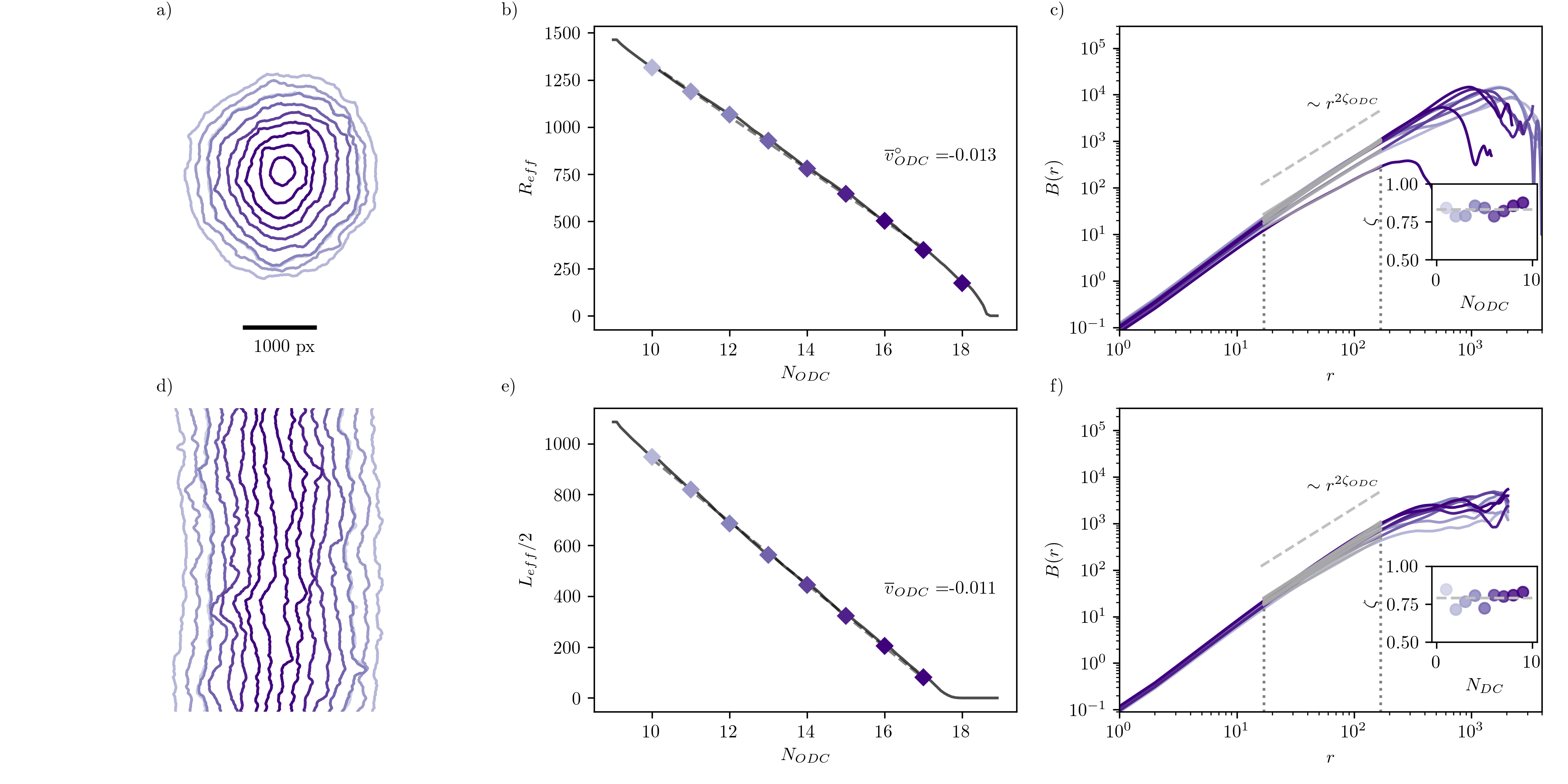}}
\end{center}
\caption{\textit{Domain shrinking under the application of square field pulses of field $-H_0$ (ODC \nccom{pulses})}: Under the application of ODC \nccom{pulses} domains shrink with larger velocities compared with the case of expanding domains subject to DC \nccom{pulses} of field $H_0$. The obtained roughness exponents are $\zeta^\circ_{ODC}=0.83\pm0.06$ and $\zeta_{ODC}=0.79\pm0.04$ for the bubble and stripe domains, respectively.}
\label{fig:ODC}
\end{figure*}

The observation of larger velocities for shrinking domains compared to expanding ones motivates a study of the DC protocol under a field $-H_0$. In figure B1 we show the domain area evolution under such a protocol, called ODC  -where "O" stands for opposite-\nccom{. This protocol was applied to a system that first evolved under the DC protocol described on the main text}. In the case of the bubble domain we obtain a larger velocity for a shrinking domain ($\overline{v}^\circ_{ODC}=-0.013$) compared to the expanding one ($\overline{v}^\circ_{DC}=0.010$). In the case of the stripe domain we also observe a difference between the DC ($\overline{v}_{DC}=0.010$) and the ODC protocols ($\overline{v}_{ODC}=-0.011$). The largest difference observed for the bubble case might be due to the curvature of the domain that induces a field $-c/R$, which will be increasing while the domain area is reduced. In the stripe case we can argue that disorder induces local curvatures that generate local effective fields. These local fields produce overall the same effect observed in the bubble case, and as a result the shrinking velocity is higher than the expanding velocity.
\\

\section{\nccom{Domain area change: The stripe case}}
\label{app:flat}

\begin{figure*}
\begin{center}
{\includegraphics[width=1\linewidth]{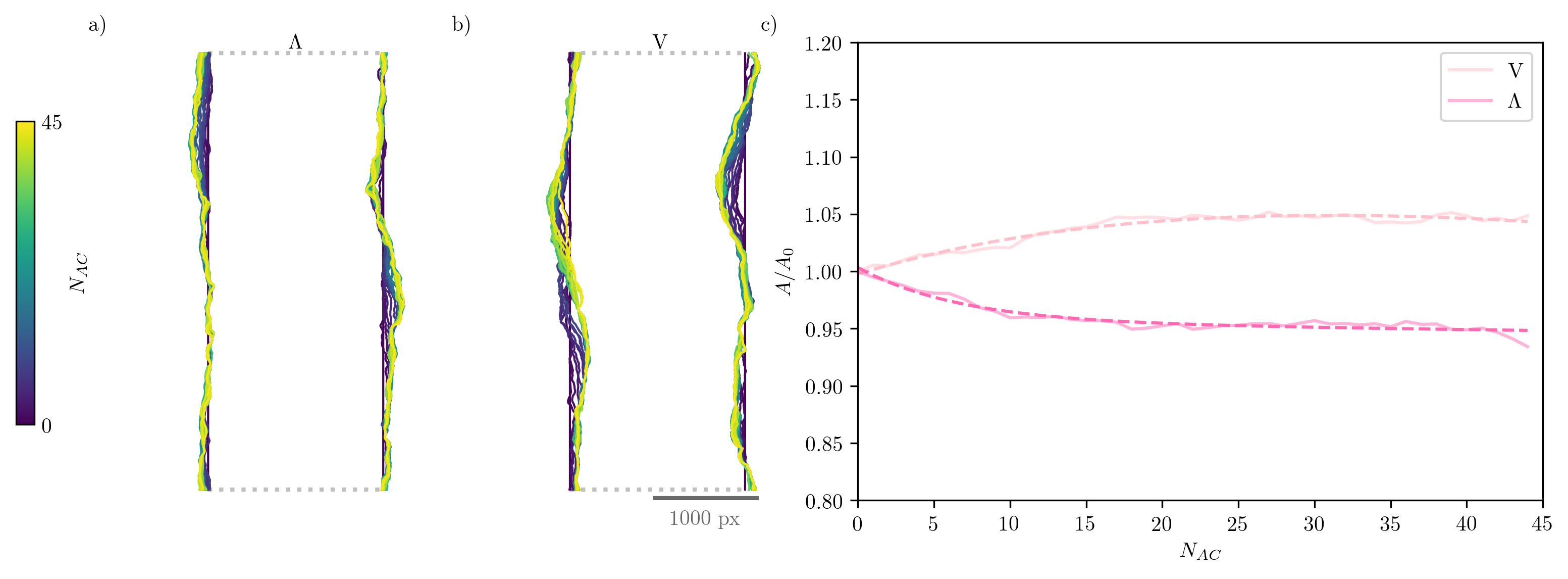}}
\end{center}
\caption{\nccom{\textit{Domain area change in the stripe case}: By initializing the system in a flat stripe state we apply two types of AC cycles: ``$\Lambda$'', where the first pulse is positive favoring domain growth, and a second negative pulse favoring domain shrinking, followed by a relaxation time. A second type: ``V'', where the first pulse is negative and the second one is positive, followed by a relaxation time. In a) and b) we show the interfaces of the stripe domain under the $\Lambda$ and V AC protocols, respectively. In c) the area loss or gain is shown with respect to the same initial state.}}
\label{fig:flat}
\end{figure*}

\nccom{The observation of area loss in the stripe case can be understood with the following field treatment. By starting from a completely flat stripe, we apply two types of $N_{AC}=45$ cycles. ``$\Lambda$'', where a first pulse is positive favoring domain growth and a second negative one favoring domain shrinking followed by a relaxation time, as shown in figure~\ref{fig:pulsesprotocol}. ``V'', in which we apply each field pulse inversely ($-H_0$ and then $H_0$), also followed by the usual relaxation time. As can be seen in figure~\ref{fig:flat}, in the $\Lambda$ case we observe a domain area reduction. However, in the V case we observe an area increase with the number of AC cycles. The domain area change is then a consequence of the state from which the field treatment starts.}

\section{Fitting of domain areas under AC cycles}
\label{app:acfit}

\begin{figure}
\begin{center}
{\includegraphics[width=1\linewidth]{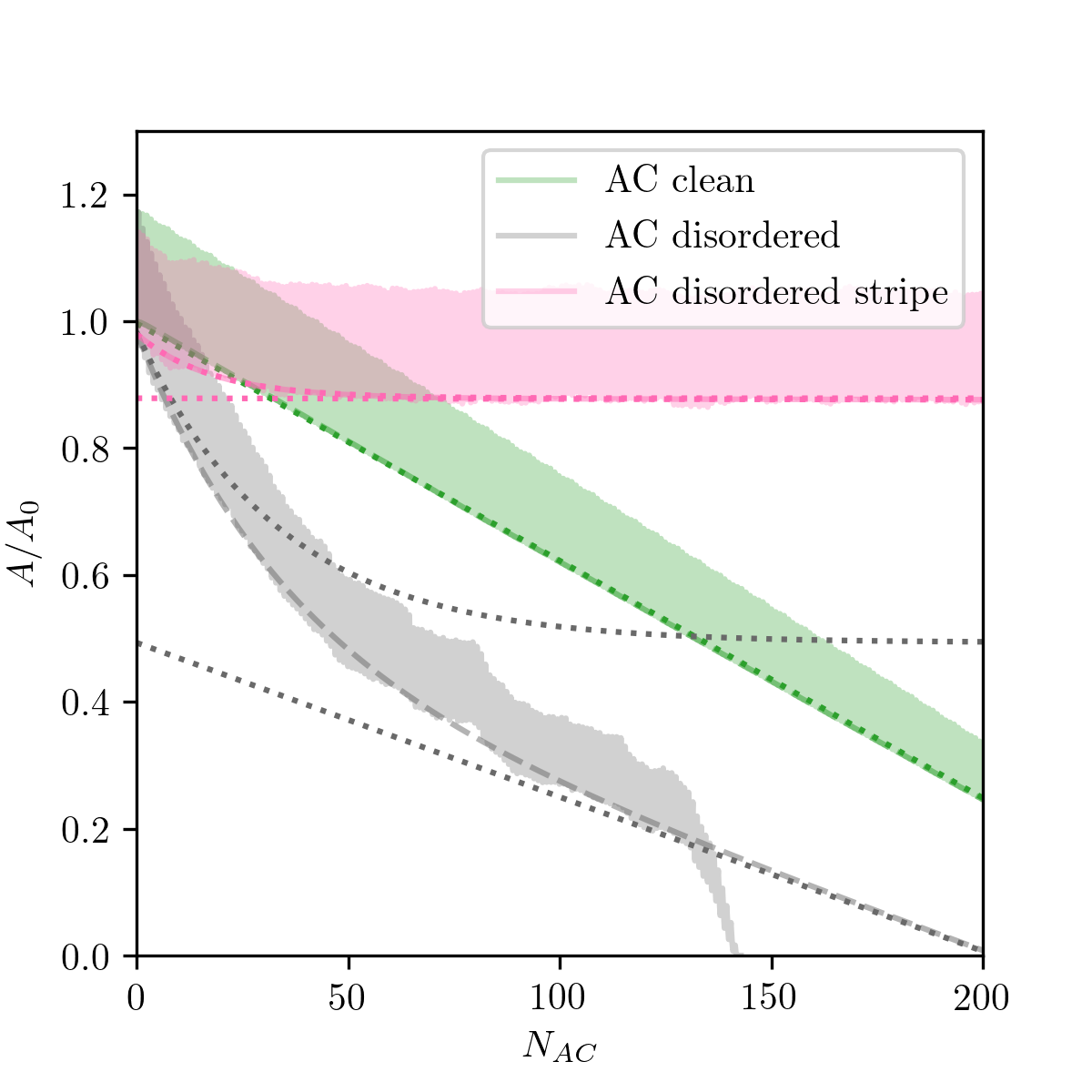}}
\end{center}
\caption{\textit{Scaling behavior of domain areas under AC cycles}: We fit the function describing the domain area after the end of each cycle with a linear combination $a_1e^{-a_2N}-b_1N+b_2$, with fitting parameters $[a_1,a_2,b_1,b_2]$. The obtained curves for each of the studied cases are shown in dashed lines. Green for a bubble domain in a clean system, gray for a bubble domain in a disordered system, and pink for a stripe domain in a disordered system. We show in dotted lines (with the same color code) the contribution of each function (linear or exponential) to the overall fit. In the stripe case the exponential dominates. In the clean bubble the linear function dominates, while in the disordered bubble the exponential dominates at short times, and the linear function dominates at larger times.}
\label{fig:fits}
\end{figure}

We fit the function describing the domain area after the end of each cycle with a linear combination $a_1e^{-a_2N}-b_1N+b_2$, with fitting parameters $[a_1,a_2,b_1,b_2]$. The obtained fitting parameters for each case are $[0,0,0.004,1.00]$ (clean bubble), $[0.10,0.05,0.00,0.09]$ (disordered stripe), $[0.50,0.03,0.002,0.49]$ (disordered bubble). The obtained curves for each of the studied cases are shown in figure C1. In the stripe case the exponential dominates. In the clean bubble the linear function dominates, while in the disordered bubble the exponential dominates at short times, and the linear function dominates at larger times.
Further exploration is needed to better understand this behavior, but we can conclude that when disorder dominates, the area reduction is described by an exponential function and when the curvature dominates the area reduction is described by a linear function.

\section{Obtaining roughness exponents}
\label{app:fitting}

\begin{figure}
\begin{center}
{\includegraphics[height=0.5\textheight]{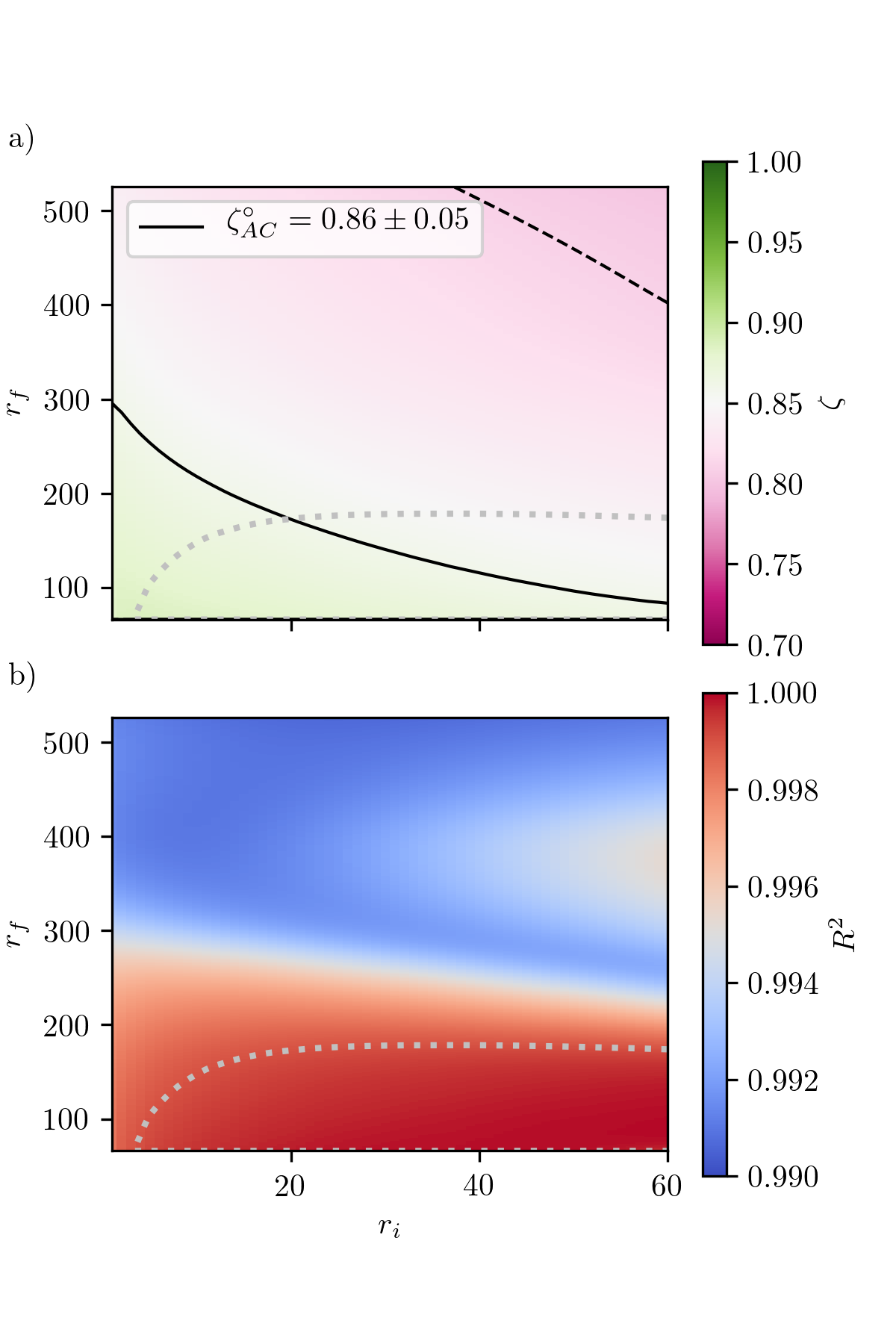}}
\end{center}
\caption{\textit{Obtaining roughness exponents}: We fix a fitting region $[r_i,r_f]$ and obtain the roughness exponent $\zeta$ and the goodness of the fit through $R^2$. In a) we show the obtained $\zeta$ for different regions for the set of functions corresponding to the bubble domain under AC cycles. In b) we show the corresponding goodness of the fit. The gray dotted line in b) delimits all the values of $R^2$ which are larger than $0.999$. This line is replicated in a). The largest fitting region with $R^2>0.999$ is of length $152$. Many sets of $[r_i,r_f]$ satisfy this condition. The one with the smallest $r_i$ is $[17,169]$. Shifts of this region by up to 6 points give the same value of the roughness exponent $\zeta_{AC}=0.86\pm0.05$. In a) the black line shows all the roughness exponents which give $0.86$. The dashed black line indicates the limit $0.86+0.05$. From this plot we can see that the selected roughness exponent is quite robust and represents the roughness exponent of many regions.}

\label{fig:R2}
\end{figure}

\begin{figure}
\begin{center}
{\includegraphics[height=0.7\textheight]{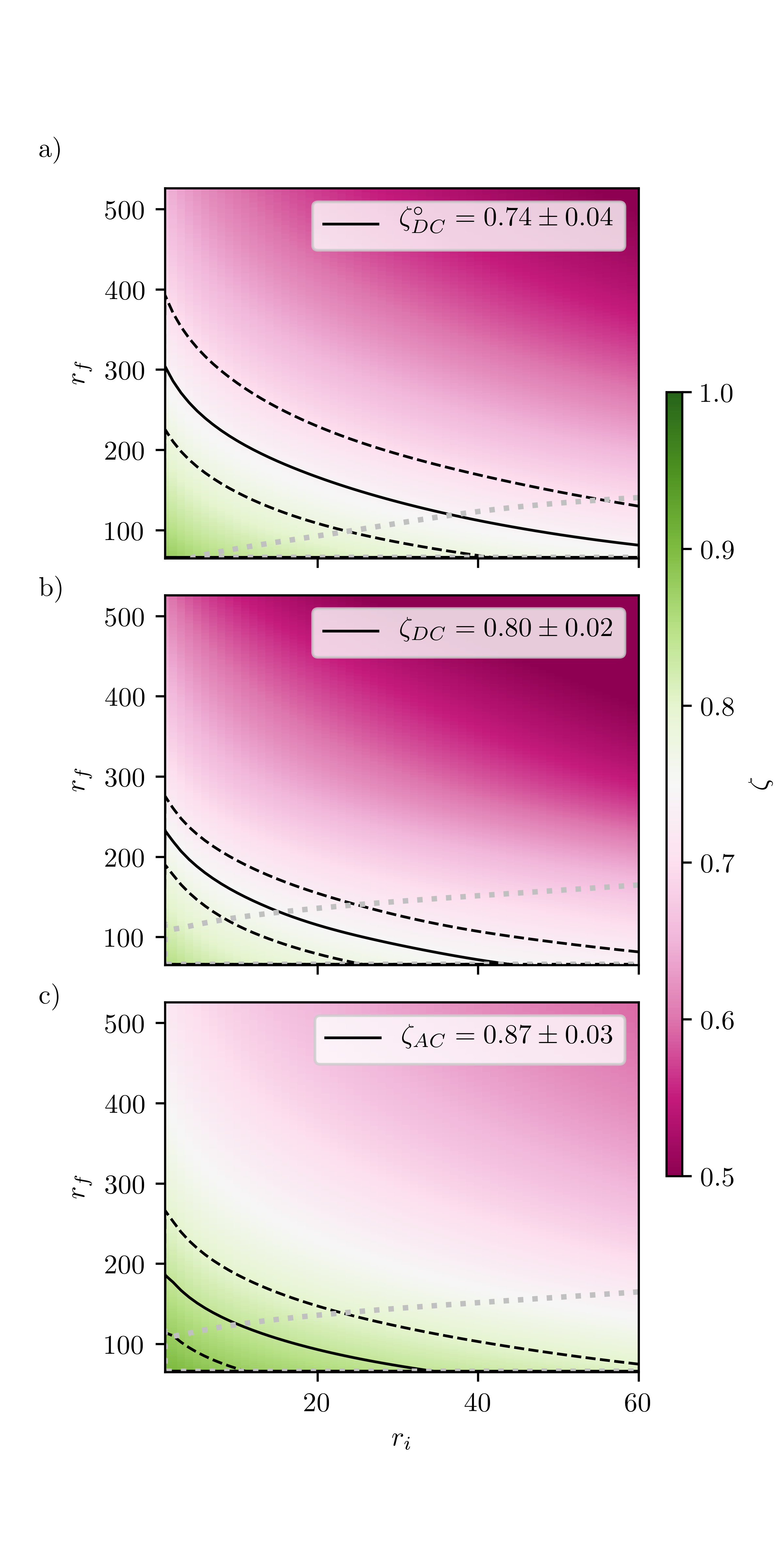}}
\end{center}
\caption{\textit{Obtaining roughness exponents for different sets}: We show the obtained values of $\zeta$ for the a) bubble DC set, b) the stripe DC set, and c) the stripe AC set. For the fitting region $[17,169]$ we obtain the roughness exponents indicated in each graph and plotted as black lines, with it respective error bars (black dashed lines). The gray dotted lines indicate the region for which the goodness of the fit $R^2$ is equal to $0.999$. All the values of $R^2$ under these curves are always larger than the threshold $0.999$.}

\label{fig:zetas}
\end{figure}

To compute the roughness exponents we have to define a fitting region. As discussed in the main text (section~\ref{sec:DomainGeometry}) this is not a trivial task. To ensure that we assign the most representative roughness exponent to each set (we have four sets defined according to the domain geometry -bubble or stripe- and according to the field protocol -AC or DC-) we use the following method.

We fit the logarithm of each function of a set with a linear function $sx+b$ for all $x$ values in the region $[r_i,r_f]$, and compute the goodness of the fit trough $R^2$. Since the roughness functions are expected to follow a power-law behavior ($B(r)=ar^{2\zeta}$), we can obtain $\zeta$ and $a$ from the two fitting parameters $s$ and $b$. For each set and each fitting region, we compute the mean values of $\zeta$ and $R^2$.

Examples of this procedure for one of the analyzed regions with $r_i=10$ and $r_f=150$ are shown in figure~\ref{fig:acdata} for the bubble case under the AC protocol. Since for cycles 36 to 85 the region in which the roughness function follows a power-law behavior is shifted due to a strong pinning center, for these cases we discard of the analysis the first 10 points.

We analyze all regions with the starting point $r_i$ ranging from 1 to 60 and $r_f$ ranging from 65 up to 526 (this limit is set by the shorter interface analyzed, that corresponds to the last observed bubble domain before the collapse). For each region $[r_i,r_f]$ we compute $\zeta$ and $R^2$. The obtained values are show in figure D1. We then scan the results and look for the largest region with $R^2>0.999$. We find that the largest region satisfying this condition is $[17,169]$. For this region we obtain $\zeta_{AC}=0.86\pm0.05$ ($R^2=0.99901$). Shifts of this region by up to six points give the same result. As shown in figure D1 a), this roughness exponent with error bars are quite representative of a big fitting region.

Since we are interested in comparing the effects of different domain geometries and different field protocols in the roughness exponent, we now analyze the roughness exponent of the other sets in the same region $[r_i=17,r_f=169]$. We obtain $\zeta^{\circ}_{DC}=0.78\pm0.04$ ($R^2=0.9991$) for the bubble under the DC protocol, while for the stripe we obtain $\zeta_{DC}=0.80\pm0.08$ ($R^2=0.999$) and $\zeta_{AC}=0.87\pm0.03$ ($R^2=0.9997$), for the DC and AC protocols respectively. It is worth noticing that if instead of fixing a region according to the AC bubble set, we follow the same method for each set, i.e, we chose the largest region for which $R^2>0.999$, for each set independently, we obtain roughness exponents which are indistinguishable from the ones obtained before. This is evidenced in figure D2.


\end{document}